\documentclass[12pt,preprint]{aastex}

\newcommand\clock{\count11=\time \divide\count11by60 \count12=\count11
\multiply\count12by-60 \advance\count12by\time
\number\count11:\ifnum\count12<10 0\fi\number\count12}

\newcommand\ov{\over}
\newcommand\E[1]{\times10^{#1}}
\def\simlt{\lower.5ex\hbox{$\; \buildrel < \over \sim \;$}}
\def\simgt{\lower.5ex\hbox{$\; \buildrel > \over \sim \;$}}

\newcommand\U[1]{{\,\rm #1}}

\newcommand\AAb{\hbox{\AA}}

\newcommand\D{\hbox{\it d}}
\newcommand\td[2]{{\mathchoice{\D#1\ov\D#2}{\D#1/\D#2}{\D#1/\D#2}{\D#1/\D#2}}}

\newcommand\al{\alpha} \newcommand\bt{\beta} \newcommand\gm{\gamma}
\newcommand\dl{\delta} \newcommand\tht{\theta} \newcommand\zt{\zeta}
\newcommand\sg{\sigma} \newcommand\Th{\Theta}
\newcommand\eps{\epsilon}
\newcommand\hsg{\hat\sigma}
\newcommand\OO{{\cal O}} \newcommand\LL{{\cal L}}
\newcommand\fperp{f_{\perp}} \newcommand\gperp{g_{\perp}}

\newcommand\NHt{N_{\hbox{\rm\footnotesize H}}}
\newcommand\NHtf{N_{\hbox{\rm\footnotesize H}\perp}}
\newcommand\NHI{N_{\hbox{\rm\footnotesize H\scriptsize I}}}
\newcommand\NHIf{N_{\hbox{\rm\footnotesize H\scriptsize I}\perp}}
\newcommand\NHIx[1]{N_{\hbox{\rm\footnotesize H\scriptsize I},#1}}
\newcommand\Nth{N_{th}}

\newcommand\HI{\hbox{H\footnotesize I}}

\newcommand\dsty{\displaystyle\strut}
\newcommand\tsty{\textstyle\strut}
\newcommand\lgt{\lg_{10}}

\newcommand\xx{\hbox to.7cm{\hss}}

\slugcomment{}

\shorttitle{Statistics of Lyman-limit and Damped systems}
\shortauthors{Bandiera and Corbelli}

\begin{document}

\title{A Comprehensive Statistical Analysis of the Gas Distribution\\
in Lyman-limit and Damped Lyman-$\al$ Absorption Systems}

\author{Rino Bandiera}
\affil{Osservatorio Astrofisico di Arcetri, Largo E.Fermi, 5, 50125 - Italy}
\email{bandiera@arcetri.astro.it}

\and

\author{Edvige Corbelli}
\affil{Osservatorio Astrofisico di Arcetri, Largo E.Fermi, 5, 50125 - Italy}
\email{edvige@arcetri.astro.it}

\begin{abstract}
In this paper we show how to use data on Lyman-limit and Damped Lyman-$\al$
absorption systems to derive the hydrogen ionization fractions and the
distribution of the face-on total gas column density.
We consider axially symmetric, randomly oriented absorbers, ionized by an
external background radiation field in order to relate the face-on total gas
distribution to that of the neutral hydrogen observed along the line of sight.
We devise a statistical procedure based on the Maximum Likelihood criterion,
that is able to treat simultaneously data coming from different surveys and
statistically recovers the ``true'' column densities in the presence of large
uncertainties:  this is especially important for Lyman-limit systems which
leave an unmeasurable residual flux at wavelengths shorter than the Lyman
break.
We make use of simulated data to look for possible observational biases and
extensively test our procedure.
For a large statistical sample of real data in the redshift range [1.75,3.25]
(collected from all published surveys) our Maximum Likelihood procedure gives a
power-law slope for the total hydrogen distribution of $-2.7$.
All together Lyman-limit systems therefore contain more gas than Damped
Lyman-$\al$ systems.
Analysis of data at other redshifts shows that more observations are needed to
reach a compelling evidence for a cosmological evolution of the slope of the
gas distribution.

\end{abstract}


\section{Introduction}

Absorption features in quasar spectra represent a powerful tool to investigate
the formation and evolution of gaseous structures in the early Universe
\citep[and references therein]{rau98}, the efficiency of processes such as
merging \citep{kau96}, gas depletion due to star formation \citep{wol95,sto00},
and ionization by the ultraviolet background radiation field
\citep{wei97,sav97}.
Also the level of cloud clustering may be investigated analyzing absorption
spectra \citep{cri97}.
Lyman-limit absorption systems (hereafter LLS) are cosmological structures
which contain enough neutral hydrogen to absorb Lyman continuum photons and
produce a break in the QSO continuum flux level.
This feature is detectable in moderate-resolution spectra whenever the opacity
to Lyman continuum photons satisfies the condition $\tau_{LL}\gtrsim1$,
equivalent to a neutral hydrogen column density $\NHI\ge1.6\E{17}\U{cm^{-2}}$.
The column density of LLS can be determined whenever a residual flux beyond the
Lyman break is detectable.
Systems with $\NHI\simgt5\E{19}\U{cm^{-2}}$ are also visible in
moderate-resolution spectra since they give rise to Damped Lyman-$\al$
absorption lines with a reference frame equivalent width $\ge5\U{\AAb}$.
In this paper we shall refer to these as Damped Lyman-$\al$ absorption systems
(hereafter DLS).
The \HI\ column density of absorbers is usually very poorly determined
observationally when $5\E{17}<\NHI<5\E{19}\U{cm^{-2}}$ since the Lyman break is
saturated, while the Lyman-$\al$ absorption line is not yet damped.
Previous studies of absorbers with poorly determined column densities have been
carried out using a coarse binning in $\NHI$.
But a coarse binning suffers by statistical problems, related to the fact that
different choices of binning may lead to different results (e.g.\ \citet{ste95}
for problems related to redshift binning), as well as by some negative
consequences on the physical description, because it does not differentiate
\HI\ column densities which correspond to different ionization conditions.
This is particularly important for $10^{17}<\NHI<10^{20}\U{cm^{-2}}$ since in
this range a smoothly varying total gas column density distribution leads to a
rapidly varying \HI\ column density distribution due to the rapid change of the
$H$ ionization fraction \citep[hereafter Paper~I]{cor00}.

To tackle this problem, we have collected data on LLS and DLS from all the
available literature and perform a comprehensive statistical analysis, based on
a Maximum Likelihood procedure, which allows a joined fit of observations with
different $\NHI$ sensitivities and uncertainties, with different ionization
conditions and at different redshifts.

The outline of this paper is the following: in Section~2 we discuss how the
observed \HI\ column density distribution is related to the total hydrogen
distribution for face-on absorbers.
A set of analytical solutions for the cross section of non-spherical absorbers,
randomly oriented in space, is reported in Appendix A.
Section~3 contains a description of the database.
Section~4 details the statistical algorithm that we use to best fit the
\HI\ column density distribution of absorbers.
Section~4 also summarizes results relative to our database.
In Section~5 we perform realistic simulations of the data, which are used to
test the effectiveness and the robustness of the statistical algorithm.

\section{Column densities relationships}

One physically meaningful quantity for investigating the formation and
evolution of gas condensations is the distribution of their gaseous mass.
However observations measure only the distribution of the neutral hydrogen
column density along the line of sight, which may present a rather complex
behavior even if the face-on total column density distribution is a power law.
In this section we show how to transform one distribution into the other.
The relationship between the total and the neutral column density has been
investigated in detail in Paper~I.
Here we focus on the effect of randomly oriented absorbers which are either
flat or of ellipsoidal shape.

\subsection{From the total to the neutral column density distribution}

We show first how a face-on total hydrogen column density distribution
$\gperp(\NHtf)$ translates into a face-on neutral hydrogen distribution
$\fperp(\NHIf)$.
For a generic relationship $\NHtf\left(\NHIf\right)$ one can write:
\begin{equation}
\fperp\left(\NHIf\right)=\gperp\left(\NHtf\right)\td{\NHtf}{\NHIf}.
\label{eq:transf}
\end{equation}
If $\gperp$ is a pure power-law:
\begin{equation}
\gperp\left(\NHtf\right)=K\NHtf^{-\al}
\label{eq:powlaw}
\end{equation}
$\fperp\left(\NHIf\right)$ is also a power law when $\NHtf=Q\NHIf^{\bt}$:
\begin{equation}
\fperp\left(\NHIf\right)=\bt KQ^{-(\al-1)}\NHIf^{-\xi},
\end{equation}
with $\xi=1+\bt(\al-1)$.
For $\bt\le1$ (an ionization fraction monotonically decreasing with $\NHtf$)
and $\al>1$, $\fperp\left(\NHIf\right)$ is always shallower than
$\gperp\left(\NHtf\right)$.

For a generic $\NHtf\left(\NHIf\right)$ relationship, $\fperp(\NHIf)$ is not
tied to a power law.
For example when there is a sharp break in the slope of
$\NHtf\left(\NHIf\right)$ at $N_b$, $\fperp$ is not continuous at $N_b$, being
larger on the side where its slope is steeper:
\begin{eqnarray}
\fperp\left(\NHIf\right)=K'
  \Big(\!\!\!&\left(\xi_1-1\right)\left(\NHIf/N_b\right)^{-\xi_1}\!\Th
         \left(N_b-\NHIf\right)&+				\nonumber\\
        &\left(\xi_2-1\right)\left(\NHIf/N_b\right)^{-\xi_2}\!\Th
         \left(\NHIf-N_b\right)&\!\!\!\Big),
\label{eq:singlebreak}
\end{eqnarray}
where $\Th(x)$ is the Heaviside step function.

Similarly we can consider multiple breaks: for example if $\NHIf=\NHtf$ for
large $\NHIf$ and $\NHIf=\eps\NHtf$ for small $\NHIf$ (as for a constant
ionization fraction), $\fperp$ follows a power law with index $-\al$ in both
regions, but the low-column density distribution is shifted by a factor
$\eps^{\al-1}$ with respect to the extrapolation of the high-column density
one.

\subsection{Orientation-averaged differential cross section}

In this section we consider the effects of the random orientation of
non-spherical absorbers on the observed column density distribution.
We derive a simple analytic solution for ``infinitely thin homogeneous slabs'',
namely those whose thickness ($2h$) is much smaller than their radial extension
($2R$), and whose face-on gas column density is constant.
We then test a simple approximation for treating ``finite thickness slabs'',
using the exact solution for homogeneous ellipsoids, given in Appendix A.

Let $\sg\!\left(\NHI,\mu\right)$ be the differential cross section of an
absorber, whose symmetry axis is tilted by an angle $\tht$ with respect to the
line of sight.
Its differential cross section, averaged over orientations, is:
\begin{equation}
\hsg\!\left(\NHI\right)={1\ov2}\int_{-1}^1{\sg\!\left(\NHI,\mu\right)\,d\mu}
\end{equation}
(where we have defined $\mu\equiv\cos\tht$).
For infinitely thin slabs of surface area $A$ and face-on column density
$\NHIf$, the averaged cross section reads:
\begin{equation}
\hsg\!\left(\NHI\right)=
{1\ov2}\int_{-1}^1{A\,|\mu|\,\dl\!\left(\NHI-\NHIf/|\mu|\right)\,d\mu}=
  {A\NHIf^2\ov\NHI^3}\,\Th\left(\NHI-\NHIf\right),
\label{eq:infslab}
\end{equation}
where $\dl(x)$ is the Dirac distribution.
Therefore the averaged cross section vanishes for $\NHI<\NHIf$, and is
proportional to $\NHI^{-3}$ for $\NHI>\NHIf$.

For absorbers of finite thickness the exact behavior of $\hsg$ depends on their
detailed shape, but the following approximate formula holds:
\begin{equation}
\hsg\!\left(\NHI\right)={A\NHIf^2\ov\NHI^3}\,\Th\left(\NHI-
  \NHIf\right)\,\Th\left(\NHIf R/h-\NHI\right).
\label{eq:approx}
\end{equation}
It is shown in Appendix A that the accuracy of Eq.~(\ref{eq:approx}) decreases
as $h/R$ grows larger.
When $h\sim R$, $\hsg$ for ellipsoidal absorbers differs noticeably from
Eq.~(\ref{eq:approx}); but, once Eq.~(\ref{eq:approx}) is convolved with
realistic $\fperp$ distributions, results keep close to the exact ones.

\subsection{From the face-on to the line of sight $\NHI$ distribution}

The line of sight $\NHI$ distribution, $f(\NHI)$, can be written as:
\begin{equation}
f\left(\NHI\right)=\int{{\hsg\!\left(\NHI\right)\ov A}\fperp\!
  \left(\NHIf\right)\,d\NHIf},
\end{equation}
which, using Eq.~(\ref{eq:approx}), gives:
\begin{equation}
f\left(\NHI\right)={1\ov\NHI^3}\int_{\NHI h/R}^{\NHI}{\NHIf^2\fperp\!
  \left(\NHIf\right)\,d\NHIf}.
\end{equation}
This equation shows that $f(\NHI)$ is sensitive to the values of $\fperp$ over
the whole range $[\NHI\,h/R,\NHI]$ (i.e.\ $[0,\NHI]$ for infinitely thin
slabs).

If there is a single break in $\fperp$, as given by Eq.~(\ref{eq:singlebreak}),
the corresponding $f(\NHI)$
is:
\begin{eqnarray}
  f\left(\NHI\right)={K\ov\NHI^3}\Bigg(\!\!\!
    &{\dsty\xi_1-1\ov\dsty3-\xi_1}N_b^{\xi_1}
      \left(\min\left(\NHI,N_b\right)^{3-\xi_1}-
      \min\left(\NHI h/R,N_b\right)^{3-\xi_1}\right)+	\nonumber\\
    &{\dsty\xi_2-1\ov\dsty3-\xi_2}N_b^{\xi_2}
      \left(\max\left(\NHI,N_b\right)^{3-\xi_2}-
      \max\left(\NHI h/R,N_b\right)^{3-\xi_2}\right)\!\!\!\Bigg).
\end{eqnarray}
For $\NHI<N_b$ and for $N_bR/h<\NHI$, $f$ follows power laws with indices
$-\xi_1$ and $-\xi_2$ respectively.
In the intermediate $\NHI$ range the behavior of $f$ is more complex: if
$\fperp$ is steeper at $\NHI<N_b$, its slope approaches $-3$ for $N_b<\NHI\ll
N_bR/h$ while it bends to connect to the power-law regime at $\NHI>N_bR/h$; in
the opposite case $f$ suddenly rises for $\NHI>N_b$, and then smoothly connects
with the high column density power-law branch.

\begin{figure}
\epsscale{.99}
\plotone{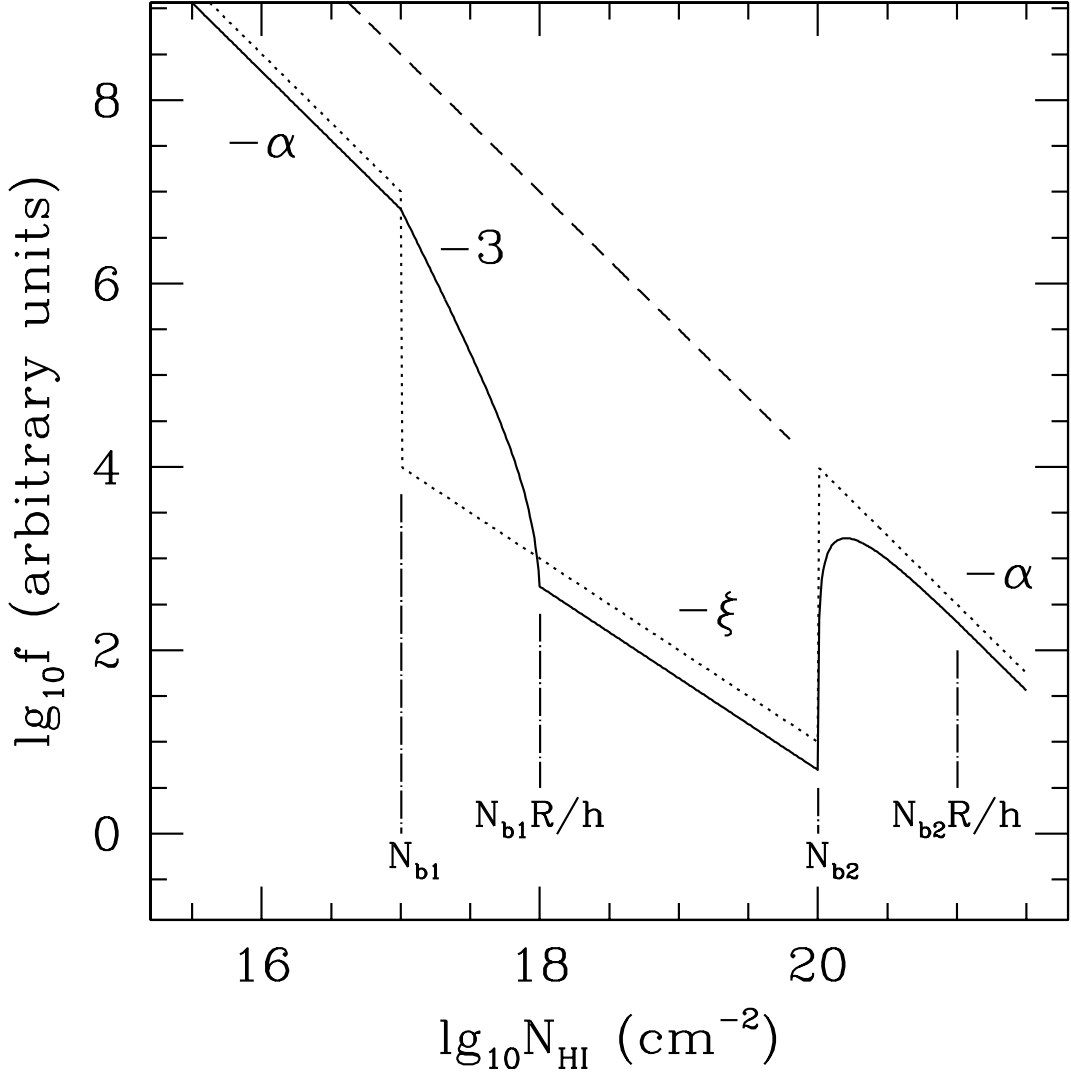}
\caption{
For a given face-on $\NHI$ distribution (dotted line) we show the projected
distribution $f(\NHI)$ (solid line).
The original face-on total gas distribution is also plotted (dashed line).
Power-law indices in regions where $f(\NHI)$ approaches a power law are
indicated.
We have used a realistic relationship between $\NHtf$ and $\NHIf$ given
explicitly in the text.
\label{fig1}}
\end{figure}

\placefigure{fig1}

Fig.~\ref{fig1} shows qualitatively the features present in $f(\NHI)$ for a
realistic $\NHtf(\NHIf)$ relationship (see Paper I): $\NHtf\propto\NHIf^\bt$
for $N_{b1}\le\NHIf\le N_{b2}$, $\NHtf\propto\NHIf$ elsewhere.
We choose $\al=1.5$, $h/R=0.1$, $N_{b1}=10^{17}\U{cm^{-2}}$,
$N_{b2}=10^{20}\U{cm^{-2}}$, and $\bt=0.001$ to emphasize the various regimes
described above.
Note that $f(\NHI)$ at low column densities is lower by a factor
$\left(N_{b1}/N_{b2}\right)^{(1-\bt)(\al-1)}$ with respect to the extrapolation
of the distribution at high column densities (dashed line).
At intermediate column densities a dip in the distribution appears, approaching
a $-3$ slope towards the low column density side.

\section{DLS\&LLS surveys and our data sample}

In this section we show which information should be extracted from the large
number of DLS\&LLS surveys available in order to build up a homogeneous
database for a global statistical approach.

In a LLS the value of $\NHI$ can be determined from the ratio between $I$, the
residual Lyman continuum flux on the blue side of the break, and $I_0$, the
unabsorbed continuum flux:
\begin{equation}
{\NHI=1.6\E{17}\tau_{LL}\U{cm^{-2}}
     =1.6\E{17}\ln\left(I_0/I\right)\U{cm^{-2}}}.
\end{equation}
If the residual flux $I$ is too small to be measured, only a lower limit to
$\NHI$ can be derived: this usually happens for $\tau_{LL}\simgt3$.
Since there are measurements sensitive to Lyman continuum optical depths as
small as 0.4, we include in our database all LLS surveys which are sensitive to
$\NHI\ge10^{16.81}\U{cm^{-2}}$.

We include in the compilation all available searches for absorption lines with
a rest frame equivalent width $W\ge5\U{\AAb}$, for which $\NHI$ can be
estimated from:
\begin{equation}
\NHI= 1.88\E{18}\,W^2\U{cm^{-2}}.
\label{eq:width}
\end{equation}

A complication comes from the fact that spurious lines with large equivalent
width may result from a blending of weaker lines with metal line systems or
Lyman-$\al$ forest lines.
Only some surveys have sufficient spectral resolution to test, by fitting a
Voigt profile, whether lines with $W\ge5\,\AAb$ are really damped.
In our database we include both the column density value determined directly
from Eq.~(\ref{eq:width}) and that derived from the Voigt profile fit, whenever
this is available.
By comparing these two values we find that there is a bias in the $\NHI$ values
estimated from the equivalent width, and derive the following statistical
correction:
\begin{equation}
\lgt\NHIx{corr}=6.261+0.705\lgt\NHIx{W}.
\label{eq:ncorr}
\end{equation}
We apply this correction to all absorption lines for which only the $W$ value
is available.
If only a lower limit to $\tau_{LL}$ can be derived from the Lyman continuum
absorption, searches for the corresponding damped Lyman-$\al$ line can be used
to establish a value or an upper limit to the \HI\ column density.
If the corresponding Lyman-$\alpha$ absorption line has not been searched for,
we take an upper limit of $5\E{21}\U{cm^{-2}}$, which is the highest measured
value of $\NHI$ in our database.

Unfortunately upper limits to saturated LLS leave large uncertainties in the
column density range $5\E{17}\simlt\NHI\simlt5\E{19}\U{cm^{-2}}$, where
theoretical predictions are very sensitive to the detailed shape of the
$\NHtf(\NHIf)$ relationship.
Hence it is essential to find a correct statistical approach for evaluating the
$\NHI$ distribution in the presence of large uncertainties to determine the
ionization conditions of the gas as well as the total gas column density
distribution in LLS and DLS.

We have collected data from the following references: \citet{tyt82,bec84,wol86,
tyt87,lan88,sar89,tur89,lan91,lwt91,cou92,bah93,sto94,lan95,ste95,wol95,sto96,
jan98,sto00}.
In order to ensure an accurate estimate of the coverage of the survey we have
excluded from the compilation the full spectrum or part of it whenever the
continuum was visually found to be very noisy or non-uniform (e.g.\ some
spectrum edges where the flux is below the 2-$\sg$ level in \citet{lan95} or
the objects 1130-106Y, PKS1206+459, MC1215+113 in \citet{bah93}).
The final number of directions i.e.\ of QSOs included is 661.
We have decided to split each direction in parts (hereafter called ``paths'')
such that each path is defined by homogeneous search parameters and contains at
most one detection.
In other words one direction may correspond to more than one path if different
sensitivities are used, or if more than one absorber is present along the line
of sight.
Our database is available in electronic form upon request to the authors.
Parameters contained in the data table can be divided into two groups: those
describing the search, and those related to the detection.

The main parameters associated with the search coverage are: {\it(i)}~the
redshift path limits, i.e.\ the upper and lower redshift for one direction in
space observed with a given sensitivity; {\it(ii)}~the sensitivity, specified
by the threshold above which $\NHI$ was detectable.

For each direction the upper redshift was set to $5000\U{km\,s^{-1}}$ less than
the QSO redshift, whenever the spectral coverage extends beyond it.
We have usually taken as the lowest redshift that given by the observers.
This does not coincide with the spectral coverage boundary in the presence of
``shadowing effect''.
The ``shadowing effect'' intervenes whenever the continuum flux, absorbed by a
LLS at $z_*$, gets below detectability: in this case the lower limit of the
path has been set to $z_*$.
If no DLS searches have been performed no further paths are appended along that
direction; otherwise, due to the different frequency of the Lyman-$\al$ line
respect to the H ionization threshold, further paths are considered between
$z_*$ and the Lyman-$\al$ shadowing redshift ($z=0.75z_*-0.25$).
The lowest redshift of the LLS search in the absence of shadowing is reported
in the database as well, since in Section~5 it will be used for simulating data
samples.
The sensitivity for a Lyman-limit search depends on the minimum opacity
detectable in the survey: typically $\tau_{LL}=0.4$, 1.0 or 1.5, corresponding
to column density thresholds $\lgt\Nth=$16.81, 17.20 and 17.38 respectively.
We have set to 0.4 the minimum detectable $\tau_{LL}$ for observations listed
in \citet{bah93} and \citet{tyt87}; we have instead set to 1.0 the threshold
for all the observations listed in \citet{lan91} and \citet{sar89} which were
revised by \citet{ste95}.
For absorption line searches with rest frame equivalent width $W \ge 5\U{\AAb}$
the sensitivity of the path was set to $\lgt\NHI=20.13$ (since we have applied
the correction given in Eq.~(\ref{eq:ncorr})).
If both LLS and DLS searches were performed on the same path the threshold of
the path was set to the lowest of the two, except in the presence of shadowing
effects.

The parameters associated with the detections are: {\it(i)}~the redshift of the
detected absorber; {\it(ii)}~the decimal logarithm of the estimated \HI\ column
density of the absorber along the line of sight ($\lgt\NHI$); {\it(iii)}~the
minimum and the maximum allowed values for $\lgt\NHI$; {\it(iv)}~the Voigt
correction factors.

When $\lgt\NHI$ could not be determined from $\tau_{LL}$ or $W$ measures, but
only limits for it are available, we have reported as $\lgt\NHI$ the arithmetic
mean between the minimum and maximum allowed values.
When instead the value of $\lgt\NHI$ could be directly determined we have
estimated the minimum and maximum allowed values by assuming a standard
spread.
In the case of non-saturated LLS we have used a spread of $\pm0.2$ in decimal
logarithm, except for the cases listed in \citet{jan98} for which we took the
uncertainties quoted by the authors.
In the case of DLS we have used again a spread of $\pm0.2$ when the line
profile has been resolved and the Voigt profile has been fitted; otherwise we
have used $\pm0.4$.
The uncertainties in $\lgt\NHI$ that we quote are usually larger than those
that can be directly derived from uncertainties on either $\tau_{LL}$ or $W$.
This is because there are other sources of errors, such as the determination of
the level of the continuum (which affects both $W$ and $\tau_{LL}$ estimates),
or the possible blending with other small absorption lines.
The Voigt correction factors are the differences between the $\lgt\NHI$ values
derived from the Voigt-profile fit and those from $W$ (Eq.~(\ref{eq:width})).

Two further columns contain flags, which are different from 0 whenever LLS and
DLS searches respectively have been performed on a given path.
A value of $-1$ implies that no systems were found, a value of 1 means that the
column density estimate is quite accurate, 2 that it is less accurate and 3
that it is very uncertain, as for DLS declared ``non damped''.
In this last case the column density value for the DLS has been set to zero, or
used as an upper limit if saturated LLS are detected at the same redshift.

\begin{figure}
\epsscale{.99}
\plotone{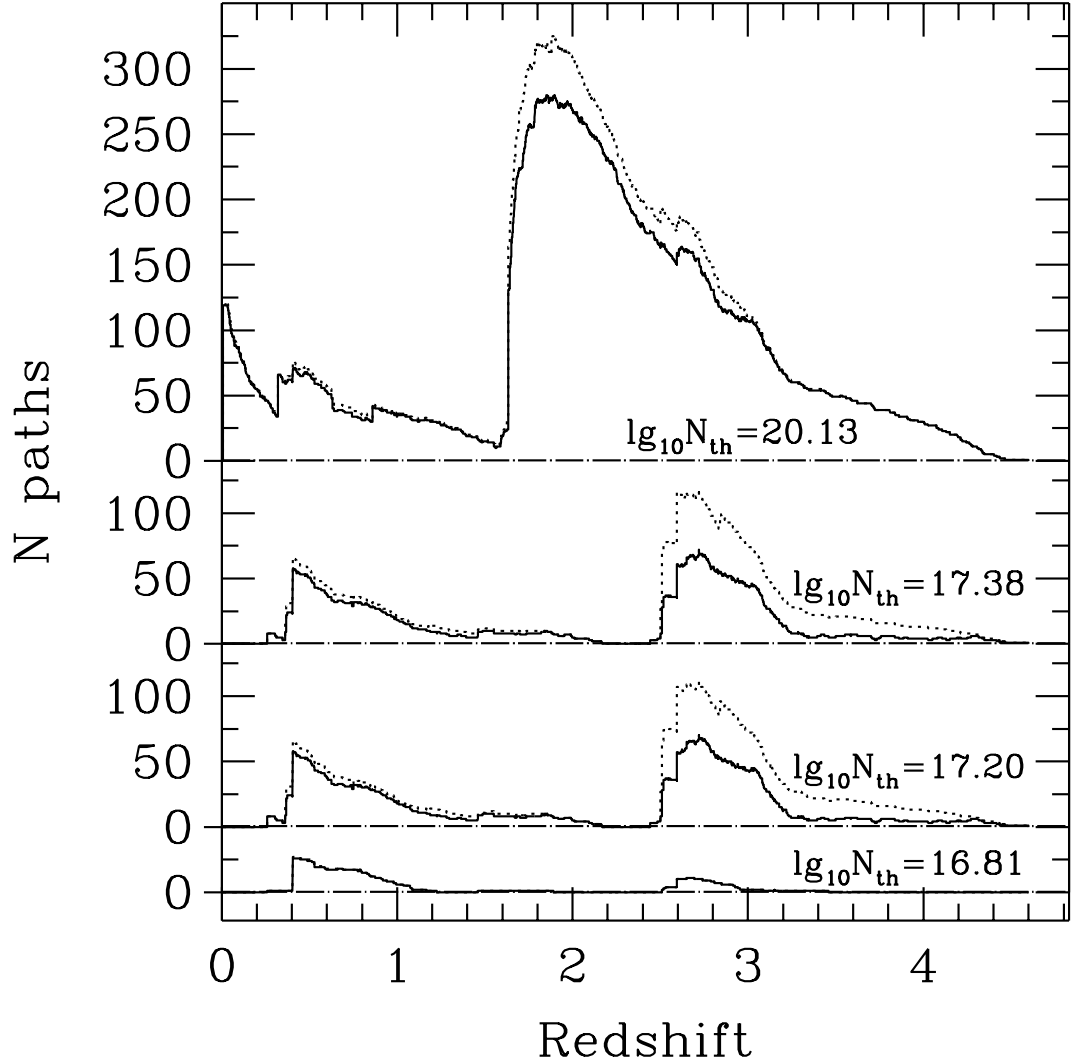}
\caption{
The redshift coverage of our sample for the various thresholds in column
density ($\lgt\Nth=$16.81, 17.20, 17.38 and 20.13).
The continuum lines refer to the actual (shadowed) coverage, while the dotted
ones refer to the case without shadowing.
\label{fig2}}
\end{figure}

The coverage of the whole sample, with and without shadowing, for various
search sensitivities is shown in Fig.~\ref{fig2}, where the inhomogeneity of
the actual coverage is evident.

\placefigure{fig2}

\section{Statistical Analysis of the Data}

Various authors have analyzed Lyman-$\alpha$ absorbers data to determine the
$\NHI$ distribution and/or its evolution with time.
Nevertheless no statistical analysis of the $\NHI$ distribution in LLS and DLS
has been attempted so far using all available data and without a pre-defined
coarse binning.
In the following we show how to perform a global unbinned statistical analysis
of heterogeneous data.

\subsection{Maximum Likelihood Analysis}

Our approach for estimating the $\NHI$ distribution is similar to the Maximum
Likelihood analysis used by \citet{sto96}, with the following two improvements:
{\it(i)}~we use a generic (not just a pure power-law) distribution for $\NHI$;
{\it(ii)}~we take into account uncertainties in the $\NHI$ measurements.
In order to simplify the notation we shall use the symbol $N$ instead of $\NHI$
in this section as well as in Section~4.2.

Let $f(N,z)$ be the distribution of column density $N$ at redshift $z$.
The average number of detections along a given direction is $f(N,z)\,\dl v$,
where $\dl v=\dl\!N\dl z$ is the area of an infinitesimal cell in the $N$--$z$
plane.
For infinitesimal cells the probability of having more than one detection in
the same cell is negligible.
The probabilities of having zero or one detection in a given cell are
respectively:
\begin{equation}
P_{0}(N,z)=\exp\left(-f\,\dl v\right);\qquad
P_{1}(N,z)=(f\dl v)\exp\left(-f\,\dl v\right).
\label{eq:probf}
\end{equation}

The likelihood for a given data set is the product of the probabilities over
``all'' cells in the $N$--$z$ plane and over all directions in space.
This product can be separated into two factors, one relative to the ``empty''
cells, and the other to the ``full'' cells:
\begin{equation}
\LL_o=\prod_{i=\hbox{\small\rm all}}P_i=
  \prod_{i=\hbox{\small\rm empty}}P_{0,i}
  \prod_{j=\hbox{\small\rm full}}P_{1,j}=
  \prod_{i=\hbox{\small\rm all}}\exp\left(f_i\dl v\right)
  \prod_{j=\hbox{\small\rm full}}\left(f_j\dl v\right).
\end{equation}
We know the total number of full cells (i.e.\ the number of detections,
hereafter $p$) but we might not be able to position the detections into
specific cells in the $N$--$z$ plane due to the large uncertainties of $N$.
For instance, if a detection has $N_j$ in the range $[N_{min,j},N_{max,j}]$,
the absorber is in one of the cells of this $N$ range, but we don't know in
which one: the probability for this event is the sum of probabilities relative
to the cells (the same considerations would apply to errors in $z$ if redshifts
are not well determined).
In the limit of small cell sizes, the sum is then approximated by an integral
of $f$ over the allowed range of $N$, and the logarithmic value of $\LL_o$ 
reads:
\begin{equation}
  \ln\LL_o=-\int{w(N,z)f(N,z)\,dNdz}+\!\!
  \sum_{j=1}^p\ln\left(
  \int_{N_{min,j}}^{N_{max,j}}{f(N,z_j)\,dN\,\dl z}\right),
\label{eq:likeli0}
\end{equation}
where $w(N,z)$ is the number of different paths at a given $z$ with a column
density threshold smaller than $N$.
The first term in Eq.~(\ref{eq:likeli0}) is related only to the parameters of
the search, and can be evaluated using the normalization condition for $f$:
\begin{equation}
\int{w(N,z)f(N,z)\,dN dz}=p.
\label{eq:normf}
\end{equation}
We can rewrite Eq.~(\ref{eq:likeli0}) using the above normalization condition:
\begin{equation}
  \ln\LL_o=p\left(\ln(\dl z)-1\right)+\!\!
  \sum_{j=1}^p\ln\left(
  \int_{N_{min,j}}^{N_{max,j}}{f(N,z_j)\,dN}\right).
\label{eq:likeli}
\end{equation}

If column densities have been measured with a rather small uncertainty for a
number $p_1$ of absorbers, Eq.~(\ref{eq:likeli}) can be approximated as:
\begin{equation}
  \ln\LL_o=p\left(\ln(\dl z)-1\right)+\!\!
  \sum_{j=1}^{p_1}\ln(\dl N_j)+\!\!
  \sum_{j=1}^{p_1}\ln\left(f(N_j,z_j)\right)+\!\!
  \sum_{j=p_1+1}^p\!\!\ln\left(
  \int_{N_{min,j}}^{N_{max,j}}{\!f(N,z_j)\,dN}\right),
\label{eq:likeliSL}
\end{equation}
which allows a faster numerical computation.
The terms with $\dl N_j$ and $\dl z$ are constant and can be neglected while
searching for the maximum value of $\LL_o$.
The limiting case $p_1=p$ reduces Eq.~(\ref{eq:likeliSL}) to the formula in
\citet{sto96}.

\subsection{The fit goodness and the true column density distribution}

The Maximum Likelihood analysis does not contain any statistical test on the
goodness of the ``best'' solution for $f$, since the absolute value of $\LL_o$
is undetermined by a constant factor.
Such a test can be done for example by comparing the theoretical versus the
observational cumulative function.
For large uncertainties of the column density values, this comparison requires
a reasonable guess at the ``true'' value of $N$ for each individual detection.
A ``reasonable'' guess does not necessarily mean that each $N$ value should
reproduce the true one but that their overall distribution cannot be
distinguished from the original one.
We will show how to find this ``reasonable'' guess in the presence of large
measurement uncertainties in $N$ in order to avoid the use of a coarse
binning.

For a given absorber $j$ the probability of having a true column density
smaller than $N$ in the allowed range $[N_{min,j},N_{max,j}]$ is:
\begin{equation}
R(N,j)=\int_{N_{min,j}}^{N}{f(N',z_j)\,dN'}\bigg/
  \int_{N_{min,j}}^{N_{max,j}}{f(N',z_j)\,dN'}.
\label{eq:rdef}
\end{equation}
$R_j$ behave as random quantities and should have a uniform distribution in the
interval $[0,1]$, independently of the shape of the distribution function $f$.

The probability of having a detection with column density smaller than $N$ in
the overall $N$ range, is:
\begin{equation}
C(N,j)=\int_{N_{th,min}}^{N}{w(N',z_j)f(N',z_j)\,dN'}\bigg/
  \int_{N_{th,min}}^{\infty}{w(N',z_j)f(N',z_j)\,dN'}.
\label{eq:cdef}
\end{equation}
The distribution of $C_j$ should also be uniform in the interval $[0,1]$, and
is equivalent to the $N$ cumulative distribution for a narrow range of
redshift.

For a sample of $x_j$ (sorted for increasing $x_j$ values, with $0\le x_j\le1$
and $1\le j\le p$) the cumulative distribution can be written as:
\begin{equation}
S(x)=\left\{	\begin{array}{ll}
			0,	& x<x_1	\\
			j/p,	& x_j\le x < x_{j+1} \\
        		1,	& x_p\le x.
		\end{array} \right.
\end{equation}
The conventional way to test whether a given a distribution is compatible with
the uniform one is to use the Kolmogorov-Smirnov test on the maximum deviation
between the two cumulative functions:
\begin{equation}
D=\max_{0\le x\le1}\left|S(x)-x\right|;
\end{equation}
the significance of $D$ being given by the function $Q_{KS}(\sqrt{p}D)$
\citep{kenstu69}.

However if the maximum discrepancy between the two distributions originates
from some bias, the Kolmogorov-Smirnov test becomes completely insensitive to
the overall match.
Such effect is reduced if instead of $D$ we use the quantity $U$:
\begin{equation}
U^2=p\int_0^1{\left(S(x)-x-\int_0^1{\left(S(x')-x'\right)\,dx'}\right)^2dx}.
\label{eq:udef}
\end{equation}
$U$ varies  with $x_j$ more smoothly than $D$ does, and its significance is
$Q_{KS}(\pi p U)$ \citep{kenstu69}.
We will use $U$ for testing the uniformity of $R_j$ and $C_j$ since it is
practically more effective and robust than $D$ to ensure a good matching over
the whole $N$ range.

For any given distribution $f$ we compute a ``best guess'' for the ``true''
column densities by maximizing the product of the significances relative to
$U_{R}$ and to $U_{C}$ (hereafter referred to as $\LL_R$ and $\LL_C$,
respectively).
There are always infinite guesses which allow a uniform distribution for
$R_j$.
However $R_j$ and $C_j$ are mutually dependent and their relationship depends
on $f$ (see Eqs.~(\ref{eq:rdef}) and (\ref{eq:cdef})).
For an incorrect choice of $f$ or of the estimated ranges
$[N_{min,j},N_{max,j}]$ it may then become impossible to derive  uniform
distributions for $R_j$ and $C_j$ at the same time.

\subsection{Numerical results}

Given an original distribution function for the face-on total column density of
absorbers of the form:
\begin{equation}
\gperp\left(\NHtf,z\right)=K(1+z)^\gm\NHtf^{-\al}
\label{eq:distrfcn}
\end{equation}
and a functional relationship between $\NHtf$ and $\NHIf$, a routine derives
the Maximum Likelihood values of $\al$ and $\gm$ for an average absorber cross
section $\hsg$ (Eq.~(\ref{eq:approx})).
$K$ is fixed by the normalization condition to the total number of detections
(Eq.~(\ref{eq:normf})); $\gm$ accounts both for cosmological effects, and for
the physical evolution of the absorbers.
Here we assume no evolution in the power-law index $\al$ in a given $z$
interval and a pure power-law dependence for $\gperp$, but it would be
straightforward to implement our routine if necessary to change these
assumptions.
Physically meaningful $\NHIf$--$\NHtf$ profiles in various redshift ranges have
been computed by simulating the ionization structure of gaseous slabs (see
Paper~I).
To identify each profile we use the parameter $X$ defined as:
\begin{equation}
X=\lgt\left(\NHtf/\NHIf\right)\qquad {\hbox{for}}\ \
  \NHI=1.6\E{17}\U{cm^{-2}}.
\label{eq:Xdef}
\end{equation}
(the corresponding $\NHIf$ depends on $\hsg$).
An increasing value of $X$ corresponds to a neutral-to-ionized transition
getting sharper and occurring at higher $\NHt$.
After determining the best $\al$ for fixed values of the parameters $X$, $\gm$
and ($h/R$) we can determine the best values for these 3 parameters by
comparing the relative Maximum Likelihood solutions.
We use the Likelihood given in Eq.~(\ref{eq:likeliSL}) with $p_1$ as the number
of detections with an $\NHI$ uncertainty, $\lgt N_{max}-\lgt
N_{min}\le\Delta\left(\lgt\NHI\right)$.
Values for $R_j$ and $C_j$ are computed for all the remaining $(p-p_1)$
detections.
Such a computation is not relevant for the Likelihood maximization itself, but
it is useful in order to compare the theoretical and the observational
cumulative.
Since it is not always possible to obtain uniformly distributed $R_j$ and $C_j$
for any set of model parameters, the uniformity level of $R_j$ and $C_j$
distributions is a further test on the quality of the fit to $f(\NHI)$
distribution.
For this reason the routine maximizes the product of $\LL_R$ and $\LL_C$.

The main results of our statistical fit procedure to data in the redshift range
[1.75,3.25], where most of the observations are, have been discussed in
Paper~I, together with the relative cosmological implications.
The Maximum Likelihood solution, $\al=2.70$ and $X=2.82$, is obtained  for
$\gamma=1.0$ and $h/R=0.2$ but $\al$ and $X$ are only weakly sensitive to
reasonable $\gamma$ and $h/R$ variations.
We now investigate the possible dependence of $\alpha$ and of the cumulative
function on the limiting value of $\Delta(\lgt\NHI)$ for the treatment of
errors, keeping $X$ fixed to 2.82.

\begin{figure}
\epsscale{.99}
\plotone{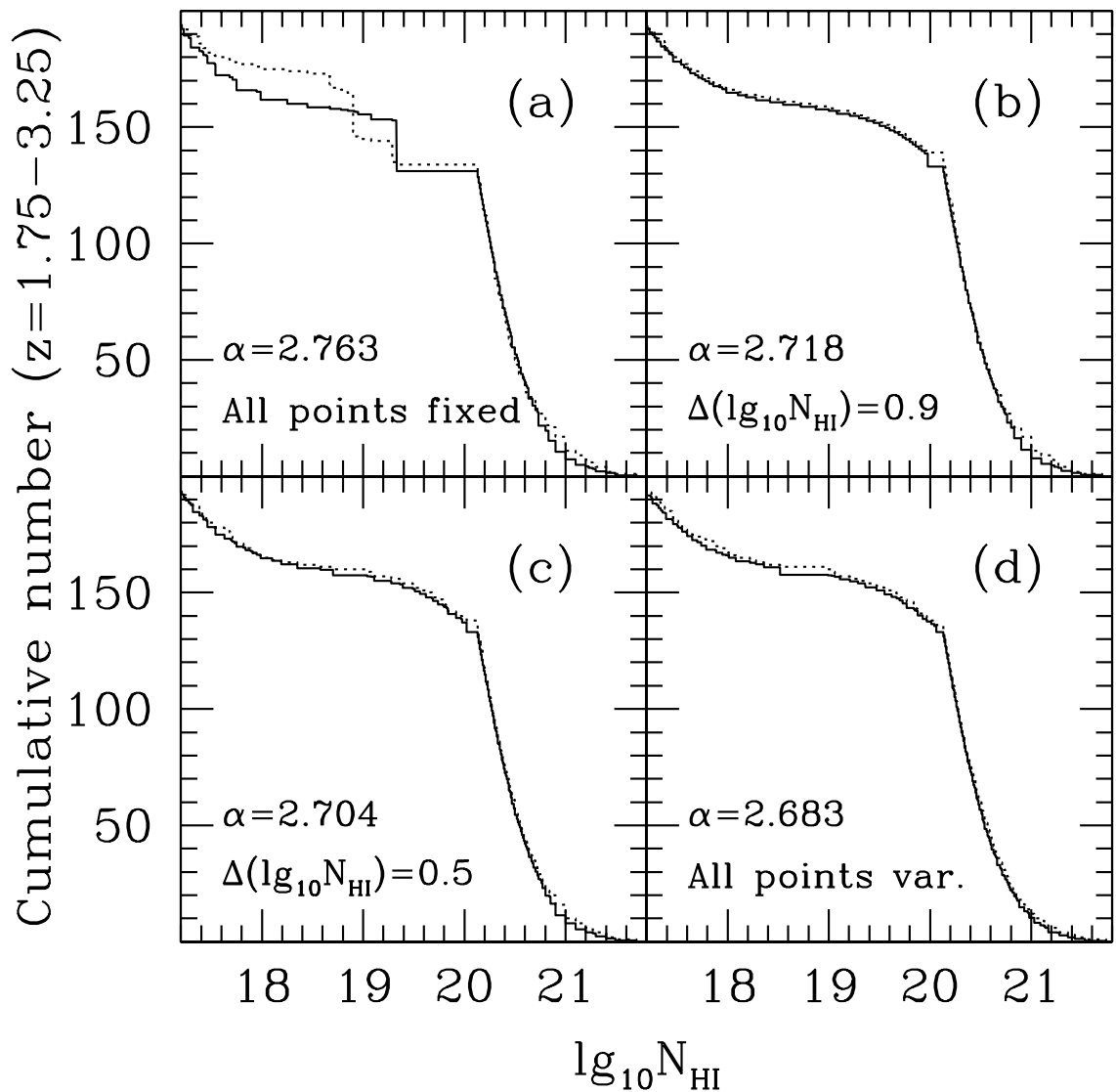}
\caption{
The Maximum Likelihood values of $\al$ and the corresponding cumulative number
of absorbers in the $z$ range [1.75,3.25] are shown for different prescriptions
of the errors treatment.
The solid line is the theoretical cumulative function, while the dotted line is
the observational one.
\label{fig3}}
\end{figure}

We consider 4 different cases:
(a) no error treatment, i.e.\ for all absorbers we use the estimated
\HI\ column density (either the measured value or the middle point in the
allowed $\lgt\NHI$ range);
(b) and (c) error treatment has been applied to all the detections with
$\Delta(\lgt\NHI)$ equal to 0.9 and 0.5 respectively;
(d) error treatment has been applied to all detections ($\Delta(\lgt\NHI)=0$).
Case (c) has been used for the results derived in Paper~I.
Fig.~\ref{fig3} shows the resulting value of $\al$ for the 4 cases and the
comparison between the theoretical and observational cumulatives.
Notice that, when the treatment of errors is operative, there is a good match
between the two cumulatives.
The ``observational'' cumulatives in these cases (dotted histograms, in
Fig.~\ref{fig3}) depend on the assumed distribution, because the $\NHI$
positions vary accordingly.

The logarithm of the Likelihood $\LL_o$, normalized to its maximum value, is
shown in Table~\ref{tab1} as we vary $\al$ for all the 4 cases.
We can see that the maximum of the Likelihood is very well determined, without
the presence of secondary maxima, and with 1-$\sg$ uncertainties of the order
of 0.1.
In Table~\ref{tab1} we also give the significances $\LL_R$ and $\LL_C$ of the
uniformity of the $R_j$ and $C_j$ distributions.
$R_j$ and $C_j$ can be uniformly distributed for a very wide range of $\al$
when the number of detections with variable $\NHI$ is large; however only near
the Maximum Likelihood value of $\alpha$ $R_j$ and $C_j$ are both uniformly
distributed.
The maximum of $\LL_C$ is better defined than that of $\LL_R$, and its position
agrees more closely with that of the maximum value of $\LL_o$.
The uniformity of the $C_j$ distribution represents a valid test of the
goodness of the Maximum Likelihood solution and of the fitting routine.

\placefigure{fig3}

\begin{figure}
\epsscale{.99}
\plotone{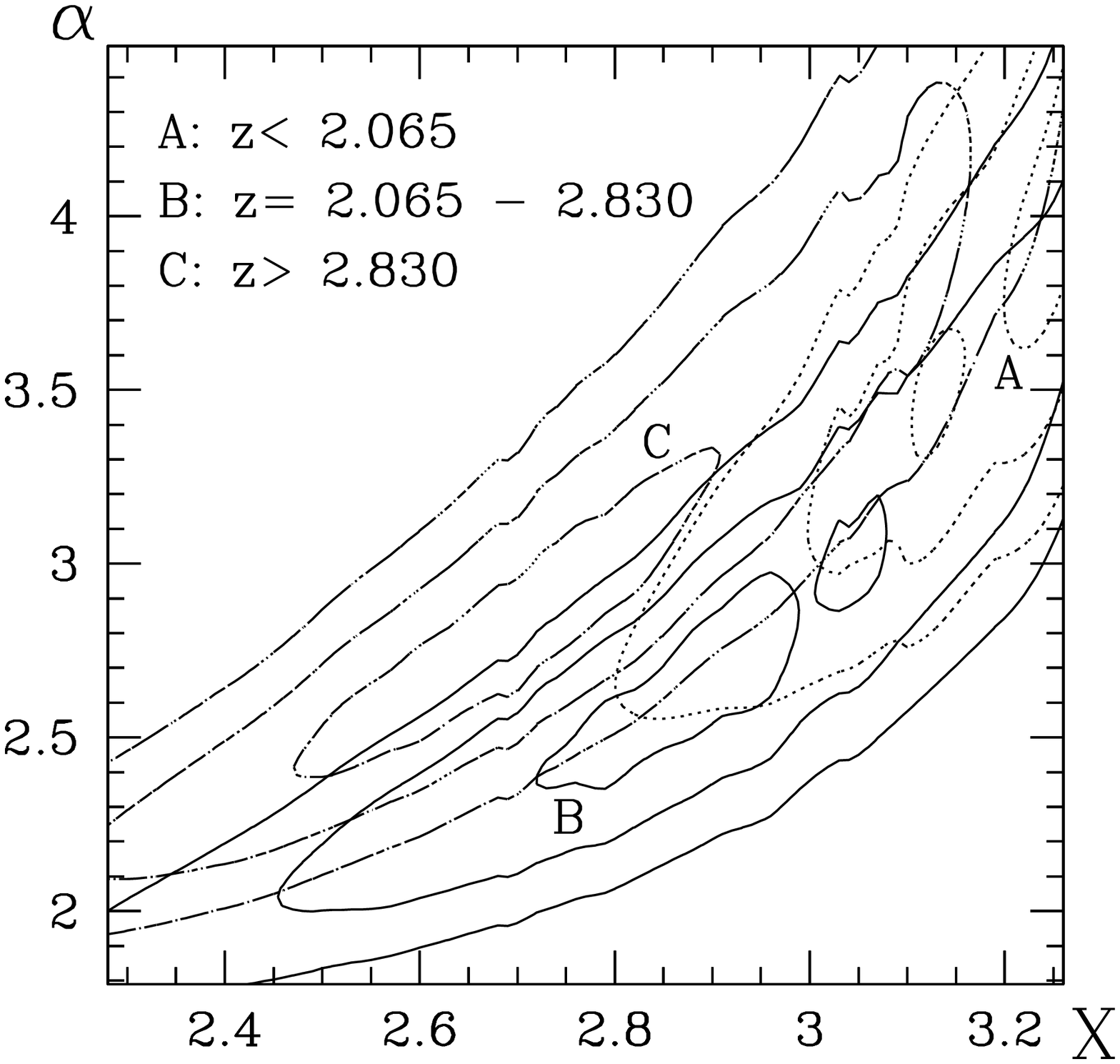}
\caption{
1-$\sg$, 2-$\sg$ and 3-$\sg$ confidence levels in the $X$-$\al$ plane, for 3
$z$-selected subsamples, labeled by $A$ (dotted lines), $B$ (solid lines) and
$C$ (dot-dashed lines).
\label{fig4}}
\end{figure}

In Paper~I we have shown that, in the $\al$-$X$ plane, the constant-probability
contours are elongated, giving a global uncertainty on $\al$ of about $\pm0.3$
when $X$ uncertainties are taken into account.
Fig.~\ref{fig4} shows the probability contours in the $\al$-$X$ plane for 3
redshift intervals which together cover all the redshifts in our compilation,
each containing the same number of detections ($\sim90$).
We find that at lower redshifts (Subsample A) the probability contours are
shifted towards higher values of $\al$ and $X$; at higher redshifts (Subsample
C) instead the allowed $\al$ values do not show significant changes, while $X$
decreases.
However, since these variations are of the same order of parameter
uncertainties, they could just be statistical fluctuations.
Further insight will come out from the simulated data analysis in the next
session.

\placefigure{fig4}

An issue that will be discussed more in detail in a future paper is whether
line blending is present in the data and could affect the above results.
The probability of blending of Lyman-$\al$ lines is relatively low for randomly
distributed absorbers, but clustering or blending with some metal lines might
imply a somewhat higher probability and affect lines with $5\le W<10\U{\AAb}$,
sometimes declared as non damped.
To address this point we have performed an analysis of the data without
considering low-resolution searches for lines with $5\le W<10\U{\AAb}$: the
$\al$ and $X$ values that we find are consistent with those relative to the
whole data sample.
Therefore we tend to believe that our data sample closely reflects the real
\HI\ distribution.

\section{Simulations}

Possible biases might affect the data analysis, either related to our
assumptions about the physical properties of the absorbers, or to our
interpretation of the observations.
In this section we show how to use simulated data for investigating the latter
type of biases.
These might derive from:
{\it(i)}~an uneven coverage of $\NHI$ and $z$ since the database contains
results from surveys with different sensitivities;
{\it(ii)}~non uniform uncertainties in $\NHI$;
{\it(iii)}~incorrect estimates of detection thresholds or of $\NHI$
uncertainties.
Simulated data can also be used to test the statistical procedure described in
the previous section, and the confidence levels properties for the best fitting
parameters.

\subsection{The numerical procedure}

Simulations supply, in a controlled way, a large number of samples to which we
can apply our statistical analysis.
We have devised a routine to simulate data in a highly realistic way:  starting
from an ``a priori'' distribution function for the absorbers total column
densities, it generates a data set compatible with a given column density and
redshift coverage.
In the present work we use a coverage consistent with the observations listed
in our database.
The basic ingredients of the simulation are:
1.~an original distribution function of $\NHtf$ for absorbers of a given
geometrical shape;
2.~an $\NHtf$--$\NHIf$ relationship for each redshift bin;
3.~a $\NHI$--$z$ coverage for the simulated survey.
After the total face-on column density has been transformed into a
line of sight \HI\ value, we introduce measurement errors similar
to those present in the actual survey compilation.
These errors are simulated by deriving first the random quantities $R_j$,
uniformly distributed, and then locating the uncertainties
$[\NHIx{min,j},\NHIx{max,j}]$ such that the position of ``true'' values $\NHI$
corresponds to $R_j$.
When $(\lgt\NHIx{min,j}+\lgt\NHIx{max,j})/2$, is above $10^{17.68}\U{cm^{-2}}$
the absorption is considered as saturated.
In this case, if there has been a search for the corresponding damped
Lyman-$\al$ but the line has not been found, the uncertainty range is set to
$[10^{17.68},10^{20.13}]\U{cm^{-2}}$; if instead the damped Lyman-$\al$ line
has not been searched for, the uncertainty range is set to
$[10^{17.68},10^{21.7}]\U{cm^{-2}}$.

For saturated absorption we simulate also the shadowing effect.
Shadowed absorbers are eliminated from the sample and the $\NHI$--$z$ coverage
of the survey is modified accordingly for LLS and DLS searches: this
``after-shadowing'' coverage is what should be compared with the real data
coverage.

\subsection{Simulation results}

\begin{figure}
\epsscale{.99}
\plotone{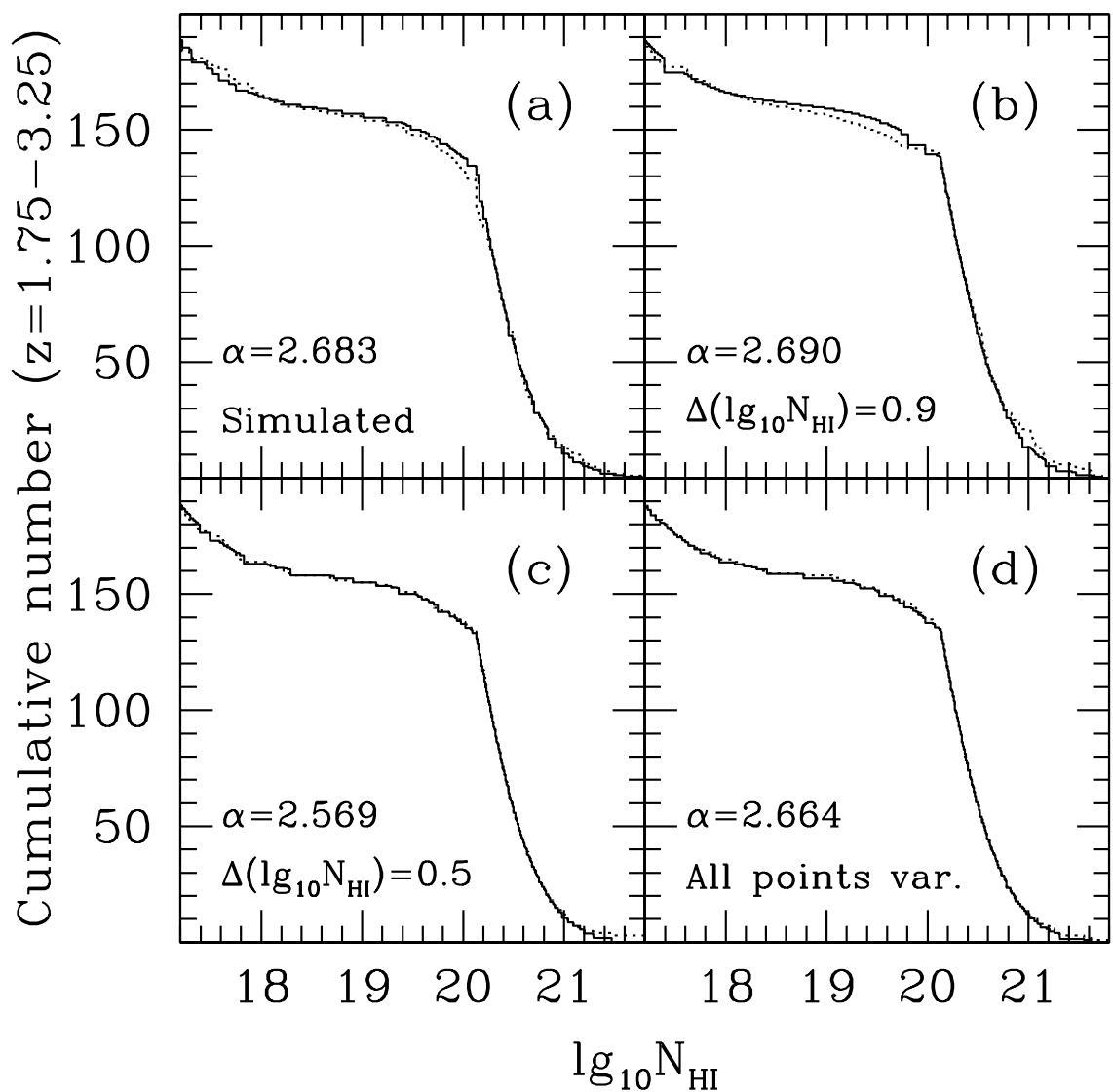}
\caption{
Various fits to the cumulative number of absorbers in the $z$ range
[1.75,3.25], on simulated data.
This figure is analogous to Fig.~\ref{fig3}, except for panel (a), which gives
the ``true'' cumulative distributions, both theoretical (solid line) and
``measured'' (dotted line).
\label{fig5}}
\end{figure}

We first simulate samples of data consistent with the Maximum Likelihood
solution $\al=2.683$, $X=2.82$, obtained for real data in the $z$ range
[1.75,3.25] for $p_1=0$.
Results of the statistical analysis on one of our simulated data samples are
shown in Fig.~\ref{fig5}, where the efficiency of our method in recovering the
original distribution is clear.
Fig.~\ref{fig5} is the analogous of Fig.~\ref{fig3} for real data, the only
difference being that now in (a) the ``true'' values of $\NHI$ for simulated
data are used, without any degradation by measurement processes and are
compared directly with the model.
The slight disagreement between the two curves in (a) is indicative of the
magnitude of statistical fluctuations in the simulation.

\placefigure{fig5}

$\LL_o$, $\LL_R$ and $\LL_C$ for the simulated sample show a behavior very
similar to that for real data (Table 1).
An increase in the number of detections on which the treatment of errors is
applied improves the match between the ``measured'' distributions and the
original ones.
Our algorithm is very efficient in recovering the original $R_j$ distribution
as well but there is no way to recover the original $R_j$ values of individual
detections.
Detections with the same range of column densities are in all respects
identical among themselves, and any permutation between their respective $R_j$
values cannot be traced (i.e.\ there is no correlation between the ``true'' and
derived $R_j$ values).

To investigate if the selection of an error threshold affects the determination
of $\al$ we have produced 5000 different simulated samples with $\al=2.683$.
On each sample we apply the analysis with the 4 different error thresholds as
in Fig.~\ref{fig5}.
The results, reported in Table~\ref{tab2}, show that the best fit $\alpha$
values are only slightly smaller than the original ones, and the dispersion for
the best $\al$ values found ($\sg(\al)$) is consistent with the average
estimated uncertainty ($<\sg>$).

\begin{figure}
\epsscale{.99}
\plotone{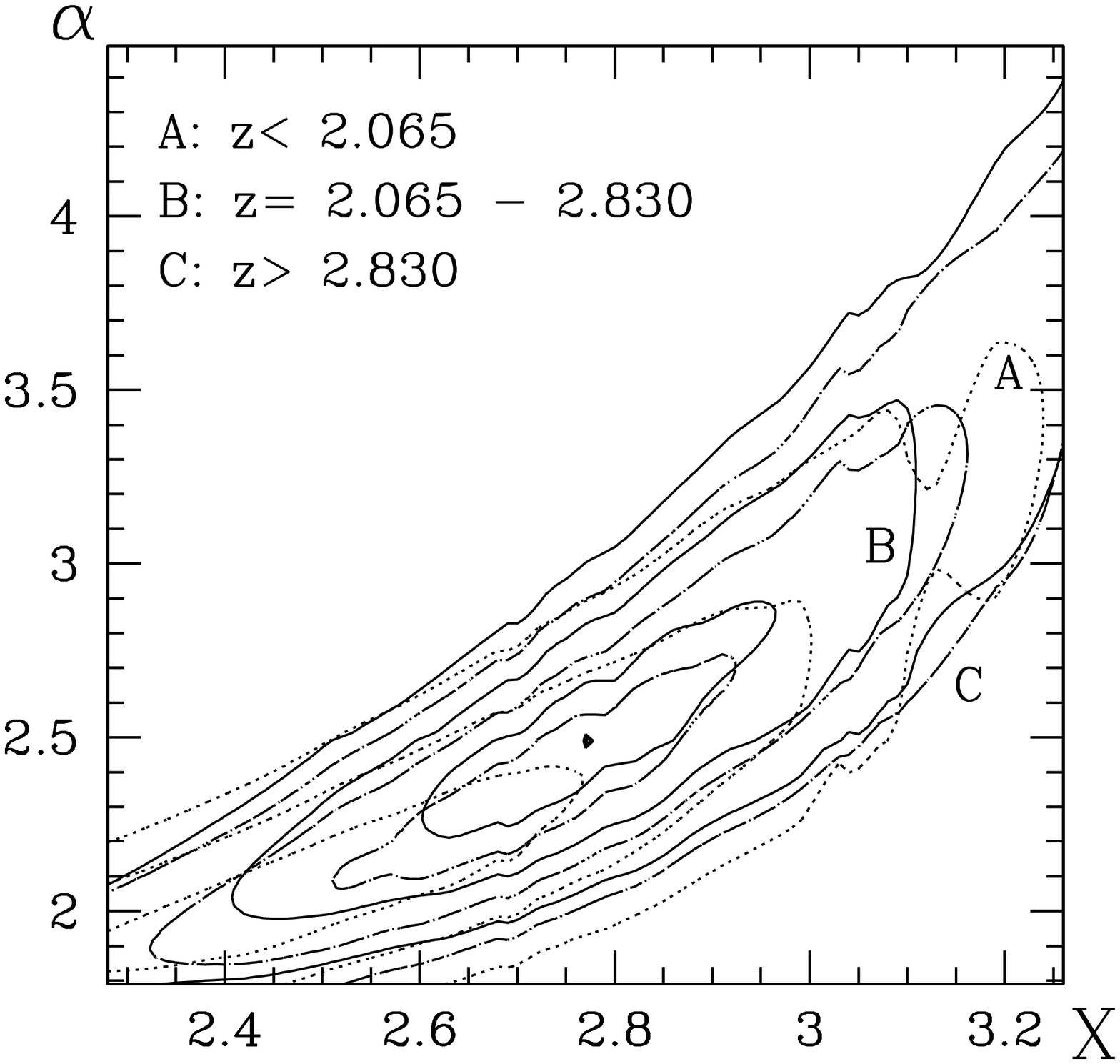}
\caption{
Map of the confidence level in the $X$-$\al$ plane, relative to a simulated
sample, for 3 $z$-selected subsamples.
All definitions in this figure are analogous to those in Fig.~\ref{fig4}.
\label{fig6}}
\end{figure}

We can also use the simulated data to check the uncertainties in the
best-fitting parameters ($\al$,$X$) by producing a confidence level map for the
simulated data, for 3 $z$-selected subsamples, analogously to what has been
done in Section~4.3 for the real data (Fig.~\ref{fig4}).
Fig.~\ref{fig6}  shows for one simulated sample that the confidence levels
present an elongation similar to the levels in Fig.~\ref{fig4}.
This means that in the real data there is no evidence of deviations of the gas
distribution from our model assumptions.
The elongation depends on a partial degeneracy of models in the $\al$-$X$
plane: $\alpha$ should increase as $X$ increases to give the same slope for the
distribution of low \HI\ column density LLS, a drawback that will be difficult
to cure by a moderate increase of the numbers of known absorbers.

\placefigure{fig6}

By looking at the three subsamples (Fig.~\ref{fig6}) it is clear that
statistical fluctuations are present mostly in the direction of the elongation,
consistently with the estimated uncertainties.
With the present sample coverage the real data are still consistent with no
evolution of $\al$ and $X$, but some evolution cannot be excluded, especially
for the high redshift sample.
For this sample in fact the shift of the probability contours is not in the
direction of elongation but mostly towards lower values of $X$.
This would imply that either the ionizing flux is lower or that the absorbing
gas has higher volume densities than at lower redshifts.
Our analysis shows that in order to have a definite answer on the evolution of
$\al$ and $X$ closer contour levels are needed.
This can be achieved by determining observationally the column density of
saturated LLS (through lines in other spectral regions or through the recover
of the quasar flux at shorter wavelengths) or by increasing the number of known
high $\NHI$ DLS.


\clearpage

\appendix

\section{Averaged cross section for homogeneous ellipsoids}
\label{app:ellips}

In this Appendix we compute analytically the differential cross section for
absorbers which can be modelled as axisymmetric ellipsoids of constant
density.
We use a cartesian coordinate system $(x,y,z)$ in which $z$ is parallel to the
line of sight.
The ellipsoid has a semi-minor axis $h$ and semi-major axes $R$ along the
orthogonal directions.
If its axis of symmetry is tilted along the $y$ axis by an angle $\tht$ with
respect to the line of sight, the surface of the ellipsoid is:
\begin{equation}
\left({\sin^2\tht\ov h^2}+{\cos^2\tht\ov R^2}\right)x^2+2\sin\tht\cos\tht\left(
  {1\ov h^2}-{1\ov R^2}\right)xz+\left({\cos^2\tht\ov h^2}+{\sin^2\tht\ov R^2}
  \right)z^2=1-{y^2\ov R^2}.
\end{equation}
The spatial extent of the ellipsoid in the $z$ direction is $2\zeta$.
Defining:
\begin{equation}
\xi^2\equiv\cos^2\tht+\sin^2\tht{h^2\ov R^2},
\end{equation}
we can write:
\begin{equation}
\zeta^2={h^2\ov\xi^2}\left(1-{x^2\ov\xi^2R^2}-{y^2\ov R^2}\right).
\label{eq:zitadef}
\end{equation}
The column density along the line of sight is $\NHI=2n\zt$.
Since $\zeta\le h/\xi$ the $\NHI$ distribution for a fixed orientation presents
an upper cut of $\NHIf/\xi$ (where $\NHIf=2nh$ is the maximum face-on column
density).

\begin{figure}
\epsscale{.99}
\plotone{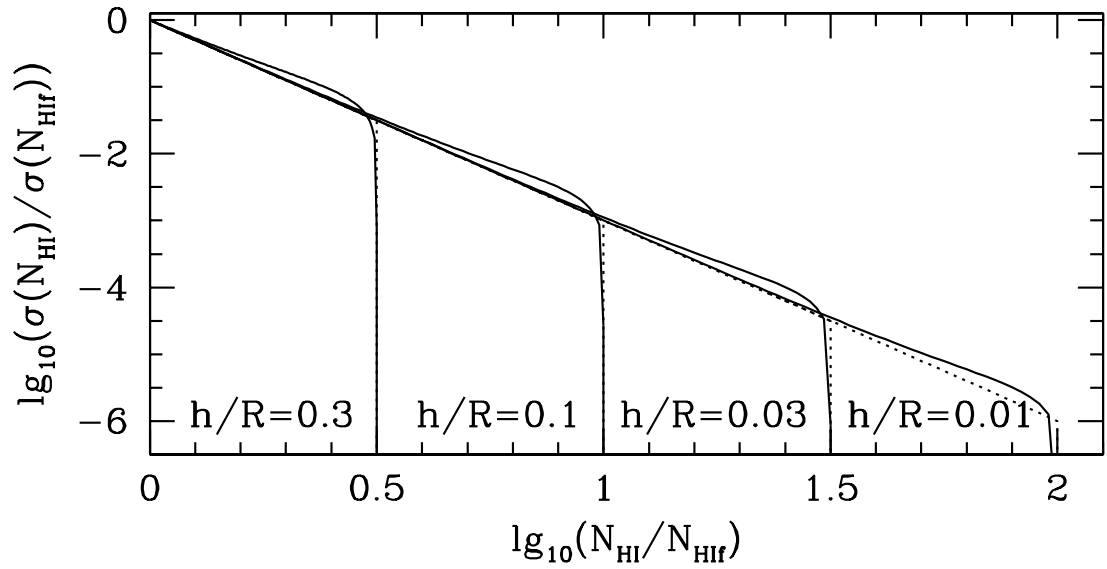}
\caption{
Comparison between orientation-averaged cross sections for ellipsoids (solid
curves) and those for the finite slab approximation (dotted curves), for
various axial ratios ($h/R=0.3,0.1,0.03,0.01$).
\label{figA1}}
\end{figure}

The ellipsoid cross section for column densities larger than $\NHI$ is:
\begin{equation}
\Sigma(\NHI)=\pi R^2\xi\left(1-\left(\xi\NHI\ov\NHIf\right)^2\right)
\end{equation}
which gives the differential cross section for one fixed orientation in space:
\begin{equation}
\sg\!\left(\NHI,\mu\right)=\pi R^2{\xi^3\NHI\ov2n^2h^2},
\end{equation}
($\mu\equiv\cos\tht$) and an orientation-averaged cross section:
\begin{equation}
\hsg\!\left(\NHI\right)=\pi R^2\int_0^1{{\xi^3\NHI\ov2n^2h^2}\,\Th
  \left(\NHIf-\xi\NHI\right)\,d\mu}.
\end{equation}
Introducing the variables:
\begin{equation}
w={R\xi\ov h}=\sqrt{1+\left({R^2\ov h^2}-1\right)\mu^2};\qquad
W={R\ov h}\min\left(1,{\NHIf\ov \NHI}\right),
\end{equation}
the average differential cross section for randomly oriented absorbers is:
\begin{eqnarray}
  \hsg(\NHI)&=&\pi R^2{\dsty2h^3\NHI\ov\dsty\NHIf^2R^3
  \sqrt{R^2/h^2-1}}\int_1^W{{\dsty w^4\ov\dsty\sqrt{w^2-1}}\,dw}
							\nonumber\\
  &=&\pi R^2{\dsty3h^4\NHI
  \left(W\left(1+{2\ov3}W^2\right)\sqrt{W^2-1}+
  \lg\left(W+\sqrt{W^2-1}\right)\right)\ov\dsty\NHIf^2R^4\sqrt{1-h^2/R^2}}.
\end{eqnarray}
For $\NHI<2nh$ the $\hsg(\NHI)$ increases linearly with $\NHI$ while it
vanishes for $\NHI>2nR$.
In the intermediate range the behavior of $\hsg$ is more complex, but for
$\NHI\ll2nR$, the following approximation holds:
\begin{equation}
\hsg\!\left(\NHI\right)={A\NHIf^2\ov\NHI^3}\left(1+\OO\left(\left(
  h\NHI\ov R\NHIf\right)^2\right)\right),
\label{eq:sgapprox}
\end{equation}
where $A=\pi R^2/2\sqrt{1-h^2/R^2}$.
Notice the similarity between Eq.~(\ref{eq:sgapprox}) and
Eq.~(\ref{eq:infslab}) for infinitely thin slabs.
In Fig.~\ref{figA1} we compare the exact solution for the ellipsoids of various
axial ratios $h/R$ to the approximate formula we use in our routines
(Eq.~(\ref{eq:approx})).

\placefigure{figA1}

Although $\hsg$ depends on the detailed shape of the absorbers, the
approximation by a $\NHI^{-3}$ power law with an upper cutoff is appropriate
for ellipsoids as well as for other geometrical shapes, especially for
thickness $h/R\ll1$.
The average value of the projected column density is instead more geometry
dependent.
For ellipsoidal absorbers this is:
\begin{eqnarray}
\hat\NHI&=&{\tsty\int_0^{\NHIf R/h}{\NHI\hsg(\NHI)\,d\NHI}\ov
          \tsty\int_0^{\NHIf R/h}{\hsg(\NHI)\,d\NHI}}	\nonumber\\
&=&{\dsty4\NHIf R\sqrt{R^2-h^2}\ov\dsty3\left(R\sqrt{R^2-h^2}+
          h^2\ln\left(\left(R+\sqrt{R^2-h^2}\right)/h\right)\right)}.
\end{eqnarray}
For $h\ll R$, $\hat\NHI$ approaches $4\NHIf/3$.
This result differs from that for a slab, for which:
\begin{equation}
\hat\NHI=2\NHIf\left(1-h/R\right)\Big/\left(1-h^2/R^2\right),
\end{equation}
whose limiting value is $\hat\NHI=2\NHIf$ for infinitesimal thickness.

\clearpage

\begin{deluxetable}{rrrrrrrrrrr}
\tabletypesize{\normalsize}
\tablecolumns{11}
\tablewidth{0pt}
\tablecaption{Best fits to the data\tablenotemark{a}\label{tab1}}
\tablehead{
\multicolumn{1}{c}{}&
\multicolumn{1}{c}{Case (a)}&
\multicolumn{3}{c}{Case (b)}&
\multicolumn{3}{c}{Case (c)}&
\multicolumn{3}{c}{Case (d)}\\
\multicolumn{1}{c}{}&
\multicolumn{1}{c}{All points fixed}&
\multicolumn{3}{c}{$\Delta\left(\lgt\NHI\right)>0.9$}&
\multicolumn{3}{c}{$\Delta\left(\lgt\NHI\right)>0.5$}&
\multicolumn{3}{c}{All points var.}\\
\multicolumn{1}{r}{$\al$}&
\multicolumn{1}{r}{$\ln\LL_o$\xx}&
\colhead{$\ln\LL_o$}& \colhead{$\ln\LL_R$}& \colhead{$\ln\LL_C$}&
\colhead{$\ln\LL_o$}& \colhead{$\ln\LL_R$}& \colhead{$\ln\LL_C$}&
\colhead{$\ln\LL_o$}& \colhead{$\ln\LL_R$}& \colhead{$\ln\LL_C$}
}
\startdata
2.05&
--25.2\xx&
--21.5& --0.083& --6.218&
--19.7& --0.022& --0.784&
--18.1&   0.000&   0.000\\
2.15&
--18.1\xx&
--15.0& --0.078& --3.784&
--13.7& --0.014& --0.253&
--12.4&   0.000&   0.000\\
2.25&
--12.3\xx&
--9.91& --0.066& --1.999&
--8.90& --0.003& --0.039&
--7.94&   0.000&   0.000\\
2.35&
--7.73\xx&
--5.94& --0.034& --0.939&
--5.25&   0.000& --0.001&
--4.55&   0.000&   0.000\\
2.45&
--4.31\xx&
--3.05& --0.011& --0.311&
--2.61&   0.000&   0.000&
--2.15&   0.000&   0.000\\
2.55&
--1.93\xx&
--1.16& --0.002& --0.054&
--0.92&   0.000&   0.000&
--0.67&   0.000&   0.000\\
2.65&
--0.52\xx&
--0.18&   0.000& --0.006&
--0.11&   0.000&   0.000&
--0.03&   0.000&   0.000\\
2.75&
  0.00\xx&
--0.05&   0.000& --0.003&
--0.09&   0.000&   0.000&
--0.18&   0.000&   0.000\\
2.85&
--0.31\xx&
--0.71&   0.000& --0.006&
--0.82&   0.000&   0.000&
--1.04&   0.000&   0.000\\
2.95&
--1.39\xx&
--2.10&   0.000& --0.025&
--2.24&   0.000&   0.000&
--2.57&   0.000&   0.000\\
3.05&
--3.20\xx&
--4.18&   0.000& --0.225&
--4.31&   0.000&   0.000&
--4.73&   0.000&   0.000\\
3.15&
--5.68\xx&
--6.90&   0.000& --0.851&
--6.98&   0.000& --0.015&
--7.47&   0.000&   0.000\\
3.25&
--8.80\xx&
--10.2&   0.000& --1.857&
--10.2& --0.002& --0.138&
--10.8&   0.000& --0.008\\
3.35&
--12.5\xx&
--14.1& --0.000& --3.175&
--14.0& --0.005& --0.476&
--14.6& --0.001& --0.131\\
3.45&
--16.8\xx&
--18.5& --0.001& --4.780&
--18.3& --0.014& --1.067&
--18.8& --0.005& --0.504\\
3.55&
--21.6\xx&
--23.4& --0.001& --6.648&
--23.0& --0.032& --1.929&
--23.6& --0.010& --1.122\\
\enddata
\tablenotetext{a}{Fits for $X=2.82$, $h/R=0.2$, $\gm=1.0$,
$\lgt\Nth=17.20$ and data in the redshift range [1.75,3.25].
}

\end{deluxetable}

\clearpage

\begin{deluxetable}{crrrr}
\tabletypesize{\normalsize}
\tablecolumns{11}
\tablewidth{0pt}
\tablecaption{Results for 5000 simulations\label{tab2}}
\tablehead{
\colhead{Case}&
\colhead{$\al$}&
\colhead{$\al-\al_{true}$}&
\colhead{$\sg(\al)$}&
\colhead{$<\sg>$}
}
\startdata
(a)& 2.549& --0.134\xx& 0.116& 0.106\\
(b)& 2.468& --0.215\xx& 0.113& 0.106\\
(c)& 2.569& --0.114\xx& 0.122& 0.116\\
(d)& 2.554& --0.129\xx& 0.120& 0.116\\
\enddata

\end{deluxetable}

\end{document}